# Stainless Steel Surface Structure and Initial Oxidation at Nanometric and Atomic Scales

*Li Ma, Frédéric Wiame,* Vincent Maurice,* Philippe Marcus**

PSL Research University, CNRS - Chimie ParisTech, Institut de Recherche de Chimie Paris (IRCP), Physical Chemistry of Surfaces Group, 11 rue Pierre et Marie Curie, 75005 Paris, France.

**Abstract**

The durability of passivable metals and alloys is often limited by the stability of the surface oxide film, the passive film, providing self-protection against corrosion in aggressive environments. Improving this stability requires to develop a deeper understanding of the surface structure and initial surface reactivity at the nanometric or atomic scale. In this work we applied scanning tunneling microscopy to unravel the surface structure of a model stainless steel surface in the metallic state and its local modifications induced by initial reaction in dioxygen gas. The results show a rich and complex structure of the oxide-free surface with reconstructed atomic lattice and self-organized lines of surface vacancies at equilibrium. New insight is brought into the mechanisms of initial oxidation at steps and vacancy injection on terraces leading to Cr-rich oxide nuclei and locally Cr-depleted terraces, impacting the subsequent mechanism of chromium enrichment essential to the stability of the surface oxide.

* Corresponding authors: F.W. (email: frederic.wiame@chimie-paristech.fr); V.M. (email: vincent.maurice@chimie-paristech.fr); P.M. (email: philippe.marcus@chimie-paristech.fr).

Stainless steels (SS) constitute one of the most technologically important class of alloys. Their durable use in our industrial society relies on the self-protection against corrosion brought by a surface barrier oxide film, the passive film, only a few nanometers thick and isolating the metallic substrate from the aggressive environment. However, the local failure of the corrosion protection can lead, in the absence of self-healing of the passive film, to localized corrosion with costly consequences and environment- and even health-jeopardizing risks. Surface analytical studies of passivity of various stainless steels at macroscopic level have shown that the passive film is markedly enriched in Cr(III) oxide/hydroxide species, which is the key for the stability of the corrosion protection [1-7]. However, the pre-passivation mechanism by which the surface oxide film initially grows may produce nanoscale chemical and structural weak points at the origin of the subsequent loss of stability [6-9]. Therefore, the development of the comprehensive insight of the stability of the corrosion protection requires not only to better understand at small length and time scales the surface reactivity in the early oxidation stages, but also to characterize in detail the structure of the surface in the metallic state prior to the formation of the surface oxide.

Ultra-high vacuum (UHV) conditions provide the environmental control necessary to investigate surface structure and oxide growth mechanisms, as shown by previous studies [10-16]. Some works investigated stainless steel single crystals under UHV [17,18], but most studies adressed the chemical composition and structure at macroscopic level by non-local techniques. Few scanning tunneling microscopy (STM) experiments have been carried out on ferritic stainless steel to characterize at the nanometric scale the single-crystal surface upon annealing [19,20] but direct observation at atomic resolution has been reported neither on ferritic nor austenitic stainless steel surfaces.

Here, we discuss the very first characterization of the structure of the oxide-free stainless steel surface at the nanometric and atomic scales, and the very first modifications related to early reaction with oxygen. We applied surface analysis at high resolution using STM under UHV conditions on a (100)-oriented Fe-18Cr-13Ni single crystal (see experimental details in Supplementary Information). This material is a good model for the most common AISI 304 austenitic stainless steel.

Figure 1 shows the surface structure of our model Fe-18Cr-13Ni stainless steel surface as repeatedly observed by STM. At the nanometric scale, it is characterized by terraces of nominal (100) orientation alternating with bunches of steps of monoatomic height hereafter referred to as multi-steps (Figure 1(a)). The terrace width is about 10–20 nm. The total height of the multi-steps is of about 3–6 atomic planes (0.54–1.07 nm) (Figure 1(b)), in agreement with the reticular distance of 0.18 nm calculated from the bulk lattice parameter of the *fcc* lattice (0.359 nm). The residual misorientation, at the origin of this terrace-step topography, is 2.5±0.2° with respect to the (100) plane. On the terraces, parallel lines constituted of dark spots are mostly aligned along the [001] direction (Figure 1(a)), with an average periodicity of about 7.6 nm. These line features were reproducibly observed after sample preparation by $Ar^+$ ion sputtering followed by annealing under UHV. They always appeared lower than the terraces independently of the imaging bias conditions, meaning that they correspond to actual topographic depressions. Their apparent depth is about 50 pm (Figure 1(c)), and could slightly vary depending on the tip used for imaging. An electronic effect may contribute to the difference of apparent height if, inside the vacancy rows, the terminating structure is bulk-like and unreconstructed, i.e. (1×1), as proposed below.

Figure 1(d) shows an atomic resolution STM image on a terrace between the periodic lines. The observed lattice is periodic and forms a square structure with a lattice parameter of 0.359 nm. This structure of the topmost plane, confirmed by low energy electron diffraction (LEED), is reconstructed and denoted (√2×√2)R45° with respect to the unreconstructed structure, denoted (1×1), of the underlying (100) planes (lattice parameter of 0.254 nm). The apparent height range between the atomic protrusions is about 20 pm (Figure 1(e)), and was observed to slightly vary with tunneling conditions and tip structure. The atomic protrusions, appearing individually brighter or darker, are randomly distributed on the square lattice but some larger areas including several protrusions also present contrast difference. These variations of apparent height of the atomic protrusions are attributed to the chemical contrast between the Fe, Cr and Ni atoms present in the topmost reconstructed plane. This kind of chemical contrast has been observed by STM on various alloys [21-25]. In the present case, it seems quite difficult to obtain a simple and clear identification of the different elements at the surface due to the rather small difference of apparent height and to the complexity of sample composition. A systematic study as a function of tunneling bias may help in identifying specific surface states of the different elements, as previously done for Fe and Cr [26,27]. It was attempted in the present work but uncontrolled changes of the imaging tip impacted the data, which supports the view of the variations of contrast being dominated by electronic effects. Future density functional theory (DFT) simulations, as done for other alloys [28,29] may also help us in the interpretation of the chemical contrast. It cannot be excluded that slight variations of the actual topographic height also contribute to the observed contrast.

Figure 1(f) shows an atomic resolution image of the structure in the area marked in Figure 1(a) where two dark lines intersect. The propagation of the perpendicular line (along [010]) is stopped by the parallel line (along [001]). The atoms always match the positions of the lattice nodes on both sides of the dark lines and no dislocation is observed. Since these lines are deeper than the average height of atoms on the terraces (Figure 1 (c)) and their width is measured as 1.05±0.05 nm, which corresponds to about 3 times the reconstructed lattice parameter, they are attributed to double rows of atomic vacancies in the reconstructed topmost plane as modeled in Figure 2. This assignment is supported by the observation of some atomic density inside the vacancy lines, as shown by the bright protrusion encircled in Figure 1(f), as well as some individual missing protrusions in the reconstructed lattice consistent with single vacancies.

The periodicity of these vacancy rows can be explained thanks to the elasticity theory [30,31]. Indeed, any metallic surface is subject to a tensile stress caused by the electron density rearrangement induced by the discontinuity of the crystal structure. In order to minimize the surface total energy, which is the sum of the elastic energy due to surface stress and of the vacancy rows creation energy, the vacancy rows self-organize by forming a periodic pattern in the [010] direction (i.e. perpendicular to [001]).[†] The small width of the terraces in the [001] direction may explain the low amount of vacancy lines in the perpendicular [001] direction due to the limited stress in that direction.

Combining all information above, the structural model shown in Figure 2 is proposed. The topmost plane is reconstructed with an atomic density half that of the underlying (100) planes. It adopts a ($\sqrt{2}\times\sqrt{2}$)R45° superstructure and contains all three alloying elements. The vacancy line consists of two rows of missing atoms as the result of surface self-organization to minimize the mechanical stress. The (1×1) structure of the underlying plane is exposed in the vacancy lines. To the best of our knowledge, this is the first direct observation of the surface structure of stainless steel at the atomic scale and the first to reveal a periodic distribution of vacancy lines on a surface.

Figure 3 shows the terrace-step structure in more details and the modifications induced by initial oxidation. On the oxide-free surface (Figure 3(a)), another structure is present in the terrace regions adjacent to the lower and upper edges of the multi-steps. This edge structure (delimited in Figure 3(a)) is continuous along the multi-steps and has a width about of 1.0–1.2 nm. It appears lower than the rest of the terrace with a relative height difference depending on the tunneling conditions. As pointed by arrows in Figure 3(a), it contains vacancies that can locally follow the step edges and are no more aligned along the [001] direction. This modified terrace structure near the step edges is attributed to the presence of two-dimensional (2D) chromium nitride surface species resulting from co-segregation of chromium and nitrogen and systematically occurring during the annealing stage of surface preparation, as evidenced by *in situ* XPS upon annealing under UHV [32] and verified after preparing tens of surface samples in the same equipment. Indeed, the chemical contrast indicates a different composition of this edge structure and the very low sub-monolayer amount of chromium nitride estimated by XPS is consistent with this interpretation. It is thought that this edge structure contributes to relieve the stress at the origin of the self-organization of the vacancy lines, precluding their alignment to reach the step edge.

---

[†] F. Wiame, V. Ruffine, L. Ma, V. Maurice, P. Marcus, Self-ordering of vacancy rows at the surface of stainless steel, to be published.

Figure 3(b) shows the surface after dosing 1 L of $O_2$ at 250°C. According to XPS [32], Cr(III) oxide is the main oxide component formed in this very early reaction stage and 2D chromium nitride is partially converted into chromium oxide. The terraces and multi-steps are still visible, as well as the reconstruction of the topmost lattice and the vacancy lines on the terraces. However, new square structures are formed on the multi-steps (one is marked) with sides aligned along the <011> directions. The height of these new structures is 0.19 ± 0.01 nm which corresponds to one atomic step of the alloy substrate, indicating that each atomic step edge of the multi-steps has reacted. These square structures are assigned to nuclei of Cr(III) oxide observed to predominantly formed by XPS. $Fe^{2+/3+}$ cations, also observed by XPS, may be incorporated in the nuclei. CrN nitride is excluded since converted after initial oxidation. The <011> directions correspond to the close-packed directions of the unreconstructed (100) surface, indicating an epitaxial relationship between the oxide nuclei and the metallic surface. On the terraces (Figure 3(c)), one notices that more vacancies are present and form new patterns adjacent to the initial vacancy lines. New vacancies also appear and accumulate in the terrace borders adjacent to the step edges, in agreement with the conversion of 2D chromium nitride into chromium oxide. This injection of vacancies in the terraces can be thus seen as the evidence of Cr atoms being "pumped" in the topmost plane, both in the regions near the step edges because of their vicinity and at more remote distances, to feed the formation of the chromium oxide nuclei after migration to the step edges.

This study unravels the surface structure of austenitic stainless steel at nanometric and atomic scales and shows a rather complex structure even on the model (100)-oriented Fe-18Cr-13Ni single crystalline surface. The slight misorientation causes the topography to alternate terraces with atomic multi-steps (step bunching) after annealing. On the terraces, the topmost plane is reconstructed with an atomic density half that of the planes below, contains all three alloying elements and adopts a ($\sqrt{2}\times\sqrt{2}$)R45° superstructure. A novel finding for metal and alloy surfaces is the presence, after annealing, of self-organized nanostructures consisting of vacancy lines formed by two adjacent rows of missing atoms, aligned along the close-packed <001> directions of the reconstructed surface lattice. The measured average period of self-organization, related to relaxation of the surface stress, is 7.6 nm. Early reactivity is observed at the defect sites of reduced atomic coordination (steps) and affect the adjacent terrace regions. Cr(III) oxide nucleates at step edges and consumes the Cr atoms locally present but also "pumps" more remote Cr atoms from the topmost plane. The Cr vacancies injected in the terraces cause a local depletion of chromium.

The local nanostructures formed at the stainless steel surface, directly observed in this work, are rich and complex. As precursors of the surface reactivity, they are foreseen to play a significant role in the subsequent evolution of the surface interacting with its environment, impacting the mechanism of chromium enrichment essential to the stability of the surface oxide film that protects the surface in a chemically aggressive environment.

**Acknowledgments**

This project has received funding from the European Research Council (ERC) under the European Union's Horizon 2020 research and innovation program (ERC Advanced Grant No. 741123, Corrosion Initiation Mechanisms at the Nanometric and Atomic Scales : CIMNAS). Région Île-de-France is acknowledged for partial funding of the STM equipment. China Scholarship Council (CSC) is acknowledged for the scholarship to the first author (No. 201606380129).

**Figure captions**

*Figure 1 Surface structure at the nanometric and atomic scales of the (100)-oriented oxide-free Fe-18Cr-13Ni surface: (a) STM image of the terrace-step topography (50 nm × 50 nm, I = 3.5 nA, V = −0.5 V); (b) height profile of terraces and multi-steps along the line in (a); (c) typical height variation on the terraces across periodic parallel lines (perpendicular to [001] direction); (d) atomic resolution STM image on a terrace (the unit cell is highlighted) (3.6 nm×3.6 nm, I = 4.0 nA, V = −0.4 V); (e) height profile along the line in (d); (f) atomic resolution STM image of the highlighted region in (a) (8 nm × 8 nm, I = 4.0 nA, V = 0.5 V).*

*Figure 2 Atomic structure of the (100)-oriented Fe-18Cr-13Ni surface with a reconstructed topmost plane including a double vacancy row: (a) STM image (4.3 nm × 1.8 nm, I = 4.0 nA, V = −0.5 V) ; (b) top view atomic model of (a) with unit cells of reconstructed and bulk-terminated planes marked; (c) average height profile of region (a); (d) side view of atomic model.*

*Figure 3 STM images of the terrace-step topography of the (100)-oriented Fe-18Cr-13Ni surface before (a) and after (b,c) early gaseous oxidation at 250°C: (a) terrace structures adjacent to the lower and upper edges of the multi-steps are delimited by dotted and plain lines, respectively, and vacancies therein pointed (levelled image, I = 0.5 nA, V = −0.5 V); (b) Cr(III) oxide nuclei formed at each edge of the multi-steps are pointed (I = 0.2 nA, V = −0.7 V); (c) new vacancies injected in the terraces and forming new lines or patterns are pointed and circled, respectively (levelled image, I = 0.2 nA, V = −0.7 V).*

**Figure 1**

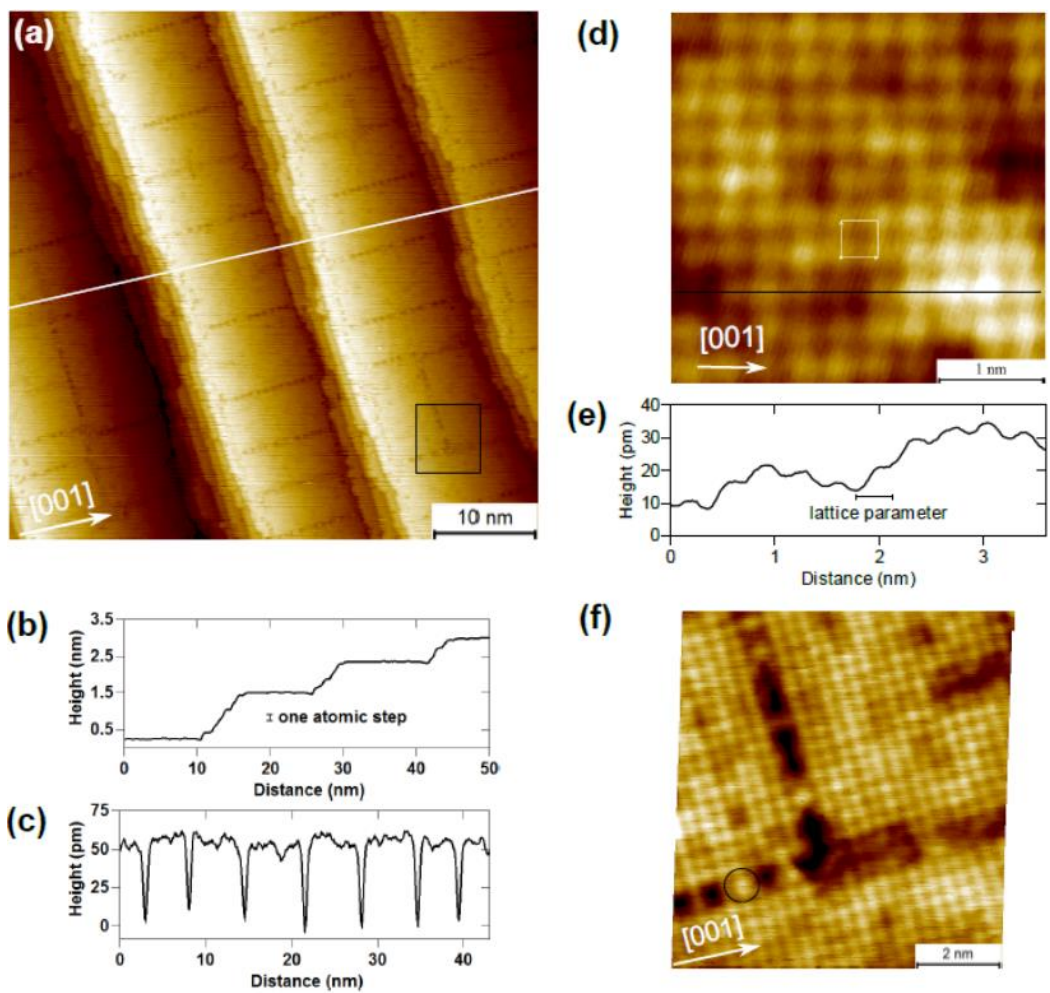

**Figure 2**

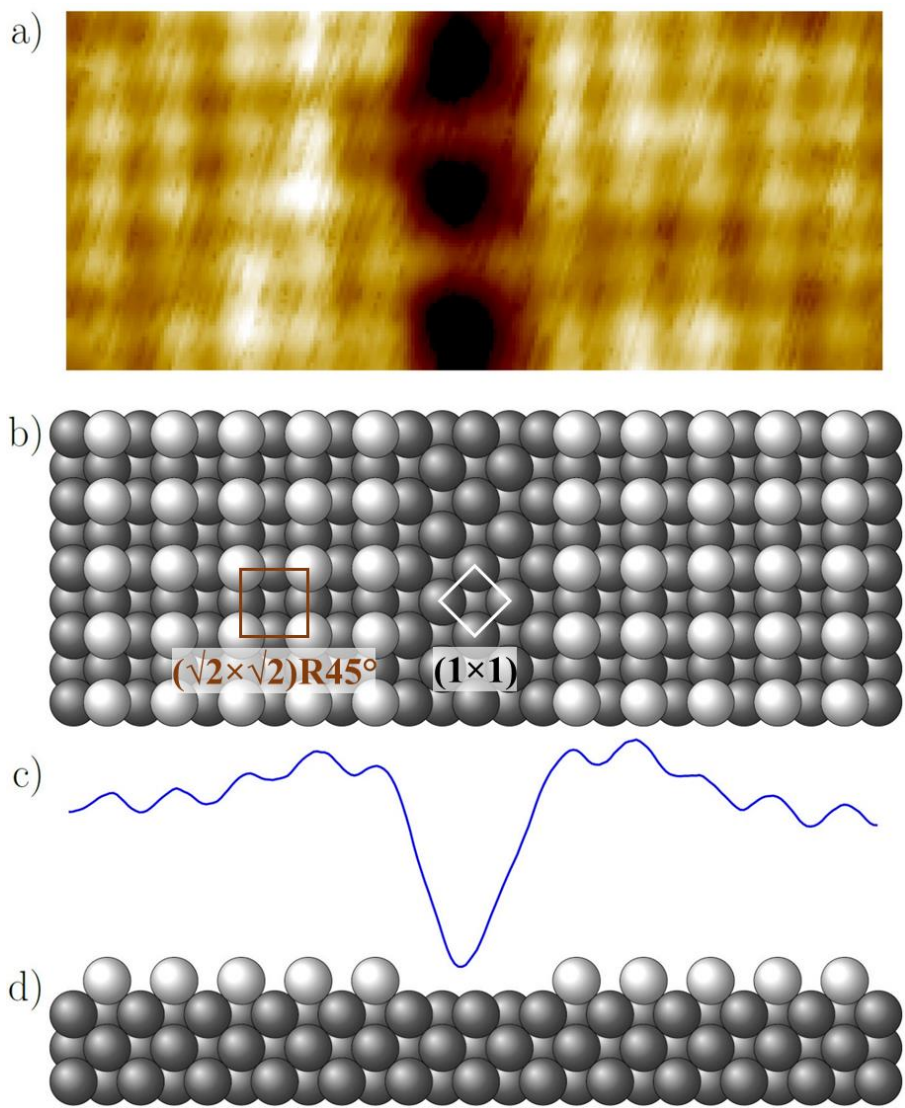

**Figure 3**

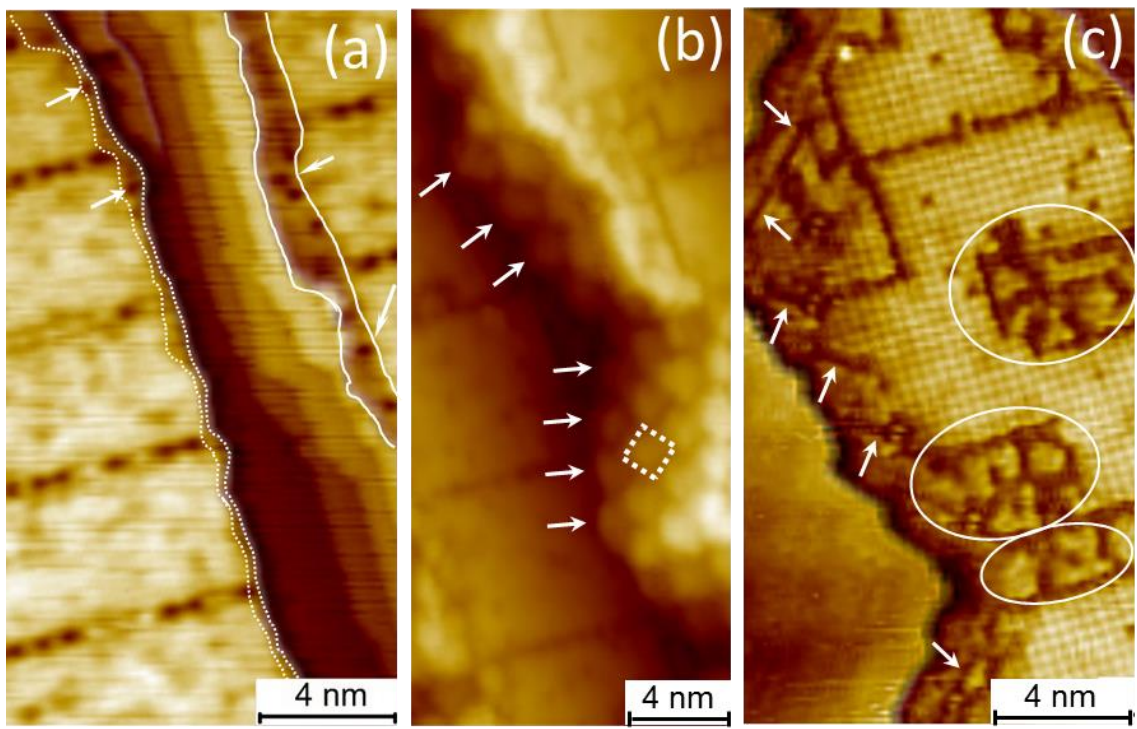

# Supplemental information to "Stainless Steel Surface Structure and Initial Oxidation at Nanometric and Atomic Scales"

Li Ma, Frédéric Wiame,[‡] Vincent Maurice,[*] Philippe Marcus[*]

PSL Research University, CNRS - Chimie ParisTech, Institut de Recherche de Chimie Paris (IRCP), Physical Chemistry of Surfaces Group, 11 rue Pierre et Marie Curie, 75005 Paris, France.

**Material and methods**

The experiments were performed in a multi-chamber UHV system from Scienta Omicron including a preparation chamber and two analysis chambers for XPS (X-ray Photoelectron Spectroscopy) and STM, maintained at a base pressure below $10^{-10}$ mbar. The preparation chamber is equipped with cold cathode ion gun for sputtering, resistive heating filaments for annealing and LEED (Low Energy Electron Diffraction) for surface periodicity characterization. The sample was heated from back side and the temperature was controlled through a thermocouple attached on the manipulator in contact with the sample.

A (100)-oriented Fe-18Cr-13Ni single crystal with 99.9999% purity was used. The sample was mechanically and electrochemically polished before introduction in the UHV system. The initial surface preparation included mechanical polishing followed by electrochemical polishing to remove the cold-work layer and was then annealed at 900°C in a flow of ultra-pure hydrogen at atmospheric pressure, as described elsewhere [17]. After introduction in the UHV system, the sample surface was cleaned *in situ* by repeated cycles of Ar$^+$ ion sputtering (1 kV, 10 µA, 10 minutes) and annealing (700°C, 10 minutes) in the preparation chamber. Surface cleanness, including the absence of oxide, and structure, reconstructed on the as-prepared oxide-free surface, were controlled by XPS and LEED, respectively. LEED was used to ascertain the crystallographic directions of the crystal. Oxygen gas ($O_2$) dosing was controlled at a pressure below $10^{-8}$ mbar in order to achieve high precision low exposure. Early oxidation was studied for 1 L exposure (1 L = $1.33\times10^{-6}$ mbar·s) at 250°C, achieved by exposing the sample in the range $1\text{-}3\times10^{-9}$ mbar for 10 minutes.

[‡] Corresponding authors: F.W. (email: frederic.wiame@chimie-paristech.fr); V.M. (email: vincent.maurice@chimie-paristech.fr); P.M. (email: philippe.marcus@chimie-paristech.fr).

STM experiments were performed with a VT STM XA. STM was calibrated from atomic resolution images on a clean Cu(111) reference surface and the tungsten tips were treated by annealing, voltage pulses and high voltage scanning. The STM measurements were carried out at room temperature. Topography leveling was performed to enhance contrast in the terraces, first by correcting the terraces from their average slope and then by subtracting the height differences between them

**Acknowledgments**

This project has received funding from the European Research Council (ERC) under the European Union's Horizon 2020 research and innovation program (ERC Advanced Grant No. 741123, Corrosion Initiation Mechanisms at the Nanometric and Atomic Scales : CIMNAS). Région Île-de-France is acknowledged for partial funding of the STM equipment. China

Scholarship Council (CSC) is acknowledged for the scholarship to the first author (No. 201606380129).